\documentclass{PoS}

\title{High-Energy Neutrino Astronomy: Current Status and Prospects.}

\ShortTitle{High-Energy Neutrino Astronomy: Current Status and Prospects.}

\author{\speaker{ Gwenha\"el de Wasseige} for the IceCube and KM3NeT collaborations
\\
        Laboratoire APC, Paris-Diderot, France, \\
        E-mail: \email{gdewasse@apc.in2p3.fr}}

\abstract{In the last decade, neutrino astronomy has taken off with two major breakthroughs, the first observation of high-energy astrophysical neutrinos in 2013 and the first evidence for gamma-rays and neutrinos from a single object published in summer 2018. In this talk, we will review these important milestones as well as the other noteworthy achievements reached by the community. We will emphasize the important role of neutrino searches in the multi-messenger era and describe the current efforts carried out in the large neutrino telescope community. We will conclude with an outlook for the coming decade.}

\FullConference{
European Physical Society Conference on High Energy Physics - EPS-HEP2019 -\\
			10-17 July, 2019\\
			Ghent, Belgium}

\begin{document}

\section{High-Energy Neutrino Astronomy}
Neutrinos are the most elusive particles in the Standard Model. Neutrinos only interact weakly, and many experiments are trying to characterize their exact nature, masses, oscillation parameters or the possible existence of a fourth neutrino. However the elusive nature of neutrinos also makes them one of the best messengers to study the non-thermal universe.
While gamma rays may be absorbed inside or outside high-energy astrophysical sources, the low interaction cross section of neutrinos allow them to escape optically thick regions and reach the Earth. Similarly, while the paths of charged cosmic rays are bent by interstellar and intergalactic fields, neutrinos are not affected because of their neutral charge. Neutrinos are thus an ideal messenger for astrophysics, provided we can establish methods to detect them despite their small interaction cross section.

The numerous neutrino telescopes around the globe provide sensitivity to neutrinos across a very large energy range. In this contribution we will focus on the large TeV neutrino telescopes, such as ANTARES and KM3NeT in the Mediterranean sea, GVD in Lake Baikal and IceCube buried in the ice of the South Pole. Since there are contributions dedicated to ANTARES~\cite{ant-eps} and GVD~\cite{gvd-eps} and the possible future Pacific Ocean Neutrino Explorer~\cite{pone-eps}, we will mainly discuss IceCube and KM3NeT.
However, all the above-mentioned detectors work on the same principle: they instrument a large volume of water or ice and detect the Cherenkov radiation emitted as a consequence of a neutrino interaction in the neighborhood of the detector. As a reference for orders of magnitude, IceCube contains 5160 digital optical modules (DOM) distributed within a volume of 1~km$^3$.
The Cherenkov technique has proven a very effective approach to detect high-energy neutrinos and has led to major breakthroughs in astroparticle physics.

\begin{figure}[t!]
	\centering
	\includegraphics[width=0.6\textwidth]{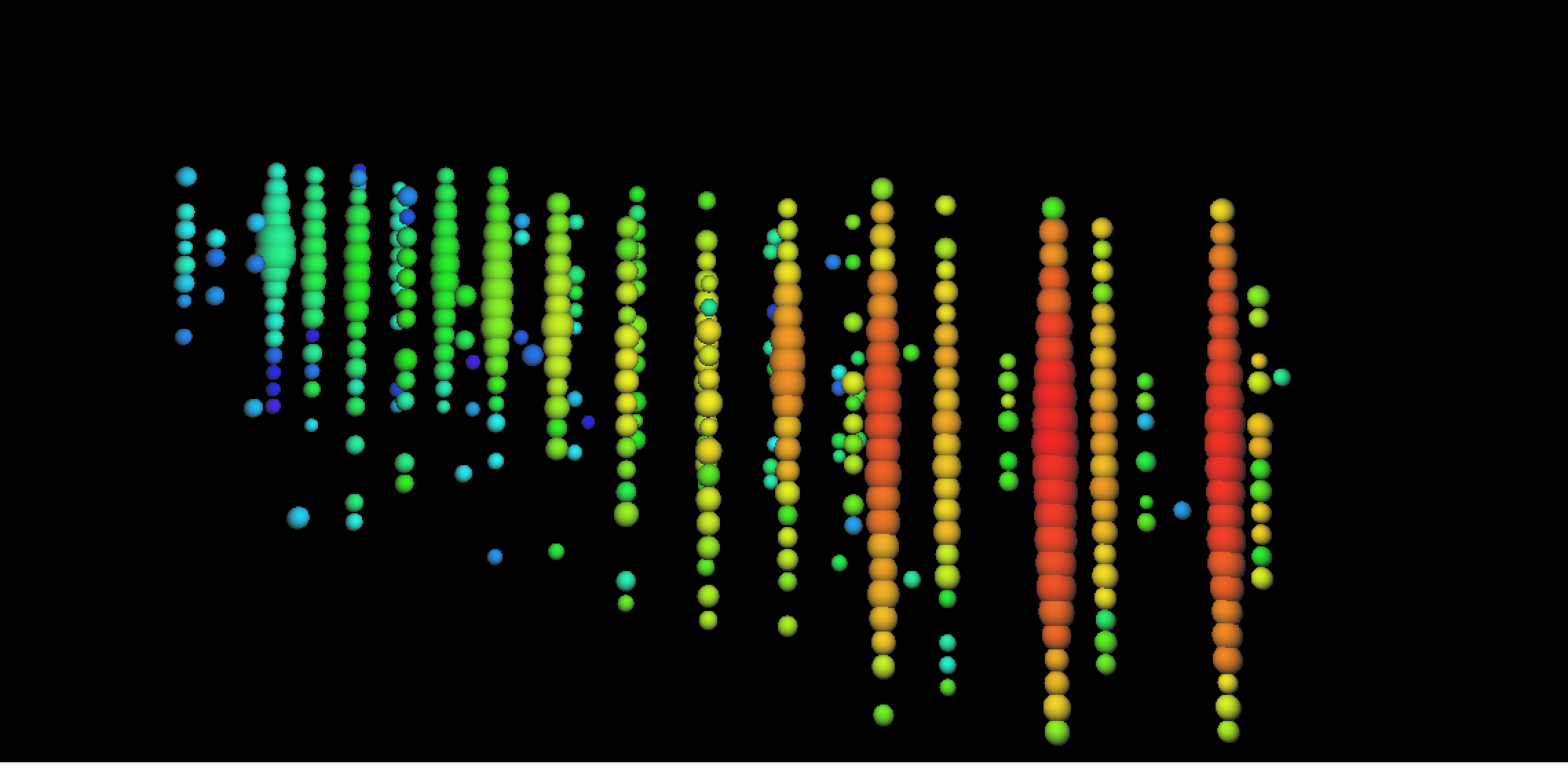}
	\caption{Neutrino event from alert IC170922A detected on September 22, 2017. The event was produced by a muon neutrino with an estimated energy of 300 TeV, pointing towards the blazar TXS 0506+056, a massive black hole with powerful jets of high-energy particles directed toward Earth. \label{fig:event}}
 \end{figure}

Fig.~\ref{fig:event} shows one of the high-energy neutrino events recorded by IceCube. Several pieces of information can be extracted using the hits recorded by the DOMs, represented by color points in Fig.~\ref{fig:event}. The number of color points and their size are proportional to the amount of light that was emitted in the ice, and directly relate to the energy of the incident neutrino. 
We can use the color code, which is an information about the timing - red meaning earlier than blue - to reconstruct the direction of the produced muon in Fig.~\ref{fig:event}, and thus infer the direction of the neutrino. We then estimate the position of the source that has produced the detected neutrino. 
Finally the spatial distribution of the hits, i.e. the topology of the event, gives us information about the flavour of the neutrino. 

As of today, only three sources of neutrinos have been identified: the Sun, more precisely neutrinos detected via the fusion reactions happening in its core~\cite{sk-sun}, the supernova SN1987A~\cite{sn} and the blazar TXS 0506+056, which is the first source detected in high-energy (TeV) neutrinos~\cite{txs}. 
This contribution presents a condensed status of neutrino astronomy, focusing on the high-energy part of the astrophysical energy spectrum, i.e. extending from GeV to PeV energies. All the neutrinos in this energy range should be produced by similar processes, proton-proton or proton-gamma interactions, and can therefore help us to identify cosmic hadronic accelerators.

\section{What did we discover?}
\begin{figure}[t!]
	\centering
	\includegraphics[width=0.6\textwidth]{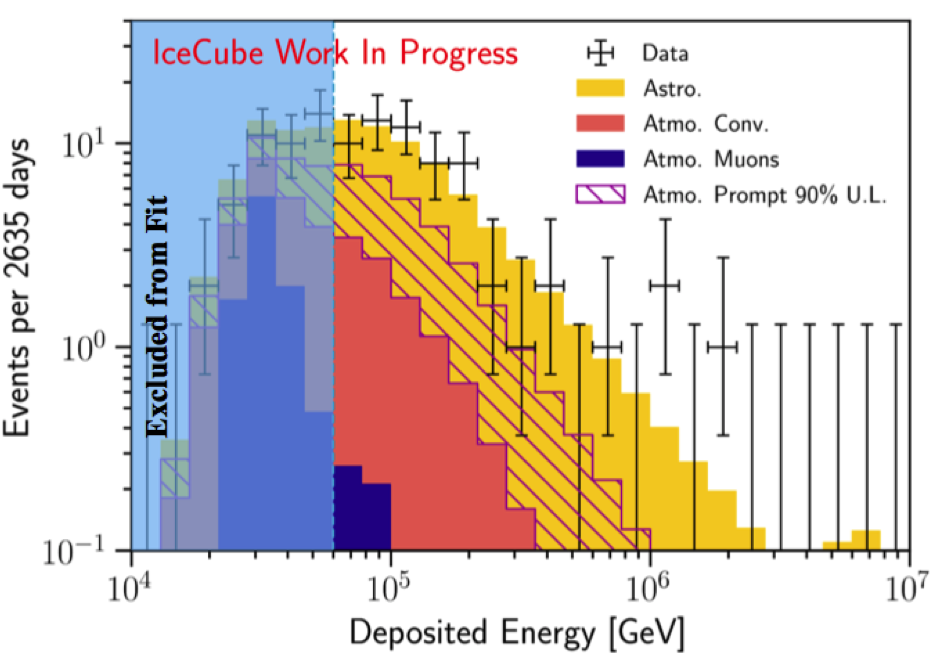}
	\caption{Distributions of observed and expected HESE events as a function of the reconstructed deposited energy. Events below 60 TeV (light blue vertical line) are not included in the fit, but one sees good data-MC agreement extending into this energy range~\cite{icrc-hese}. \label{fig:hese-spectrum}}
 \end{figure}

In 2013, IceCube announced the first detection of high-energy astrophysical neutrinos~\cite{sciencepaper-2013}. This analysis, now commonly called High-Energy Starting Events (HESE), only selects the brightest events seen in the detector, with a deposited charge > 6000 photoelectrons, and requires the event to start within the detector by using the outer DOMs to veto background events from cosmic-ray showers. These criteria lead to 75\% astrophysical purity in the HESE sample~\cite{neutrino-nancy}, making this analysis an ideal tool to measure the cosmic neutrino energy spectrum. Fig.~\ref{fig:hese-spectrum} shows the most recent update of the HESE analysis using 7.5 years of IceCube data containing 103 neutrino events with 60 events above 60 TeV~\cite{icrc-hese}. This update uses a revised calibration and ice model compared to the previous HESE data set, leading to some changes in the reconstructed direction (RA, Dec) and energy of the detected neutrinos. 
No point source has been identified yet using this data set and there is no correlation found with the Galactic Plane. The best fit energy spectrum is a single power law with spectral index $\gamma$ = 2.89$^{+0.20}_{-0.19}$  and an all-flavor flux normalization $\Phi = 6.45^{+1.46}_{-0.46}\times 10^{-18}$ GeV$^{-1}$~cm$^{-2}$~s$^{-1}$~sr$^{-1}$~\cite{icrc-hese}. We note that the fit does not prefer a broken power law model.

\begin{figure}[t!]
	\centering
	\includegraphics[width=0.95\textwidth]{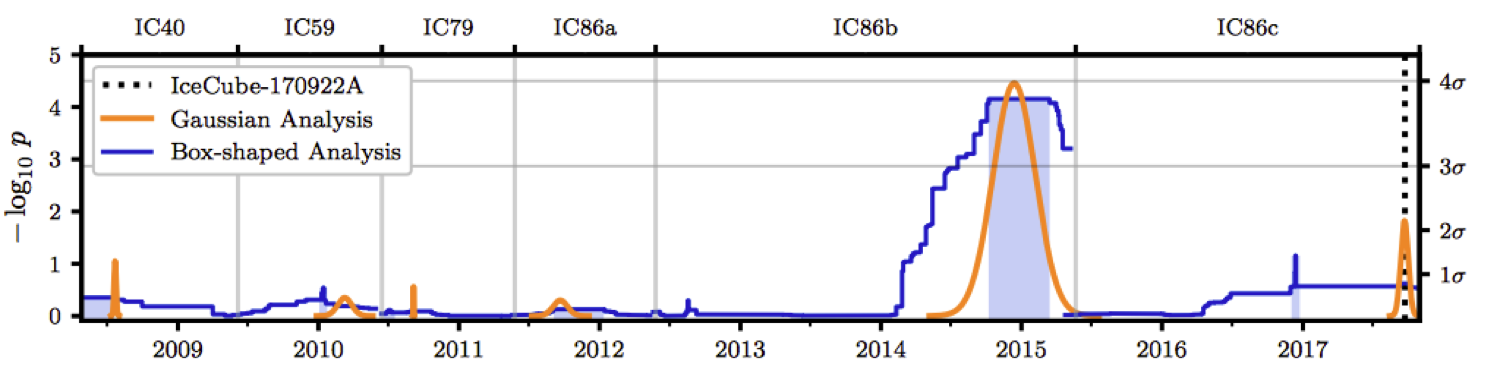}
	\caption{IceCube archival data from the region of IC-170922A, Apr 5, 2008 to Oct 31, 2017~\cite{txs}. \label{fig:timeline-txs}}
 \end{figure}

In 2018, the IceCube Collaboration, together with radio, optical, and gamma-ray telescopes, reported the first multi-messenger evidence of a flaring blazar in coincidence with the high-energy neutrino event IC-170922A shown in Fig.~\ref{fig:event}. Besides being the first common source of gamma rays and neutrinos, TXS 0506+056 is also a source candidate for the high-energy cosmic rays detected at Earth~\cite{txs}. 
A search of archival data from IceCube also revealed evidence for an earlier neutrino "flare" at the location of TXS 0506+056~\cite{txs-2}, shown in Fig.~\ref{fig:timeline-txs}. A total of 13 $\pm$ 5 neutrino excess was found in 2014-2015 over 110 days, in addition to the high-energy event detected in 2017.
The 2015 neutrino flare has by itself a significance of 3.5$\sigma$, estimated when performing identical searches using randomized event directions~\cite{txs-2}.
\section{What else, what is new? }
In addition to the previously described analyses that constitute major breakthroughs in astroparticle physics, the large neutrino telescopes have been working on numerous other analyses in order to identify neutrino sources, characterize the astrophysical neutrino spectrum, and use the astrophysical neutrino candidates to constrain neutrino physics. As examples, we refer readers to recent ICRC proceedings published by the IceCube and ANTARES collaborations~\cite{icecube-proc1,icecube-proc2,ant-proc1,ant-proc2}. 

\begin{figure}[t!]
	\centering
	\includegraphics[width=0.55\textwidth]{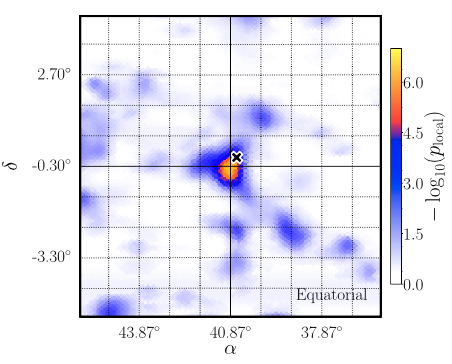}
	\caption{Local pre-trial p-value map around the most significant point in the Northern hemisphere. The black cross marks the coordinates of the galaxy NGC 1068~\cite{icrc-tessa}.\label{fig:ps10y}}
 \end{figure}
Among these new results, we highlight the 10-year all-sky point source search in IceCube~\cite{icrc-tessa}. Using the neutrino candidates detected by IceCube, the analysis performs a scan over the entire sky and evaluates the likelihood of signal over background at each location. The hottest spots are defined as the two points in the Northern and Southern hemispheres with the smallest p-values. The Northern hotspot is illustrated in Fig.~\ref{fig:ps10y}.
One can also consider correlation between the neutrino events and catalogs of sources. In this analysis a list of 110 source candidates was created from a list of gamma-bright sources that may produce neutrinos. The catalog includes active galactic nuclei, starburst galaxies, and galactic gamma ray sources. The brightest neutrino hotspot in the Northern Hemisphere coincides with the brightest object in the catalog,  NGC 1068 (M77), a Seyfert II galaxy located at a distance of 14.4 Mpc.  When the significance of NGC 1068 is compared to the most significant excesses in the Northern catalog from background trials, the post- trial significance is 2.9$\sigma$~\cite{icrc-tessa}.  The brightest source in the Southern hemisphere has a 55\% post-trial significance, and is therefore compatible with background.

Significant progress towards real-time multi-messenger detection had also been made by the large neutrino telescope community. For more information, we recommend descriptions of the real-time monitoring carried out by IceCube and ANTARES in~\cite{icrc-dawn} and~\cite{icrc-damien} respectively.
The multi-messenger connection works in two directions: the neutrino telescopes send alerts to electromagnetic (EM) partners when interesting neutrino candidate events are detected, and they also respond to alerts sent by EM or gravitational wave partners and search for a potential neutrino counterpart. We note that the alert categories from IceCube have recently changed to Gold (1 expected alert/month) and Bronze (1.3 expected alerts/month) alerts with a 50\% and 30\% signal probability respectively.

\begin{figure}[t!]
	\centering
	\includegraphics[width=0.9\textwidth]{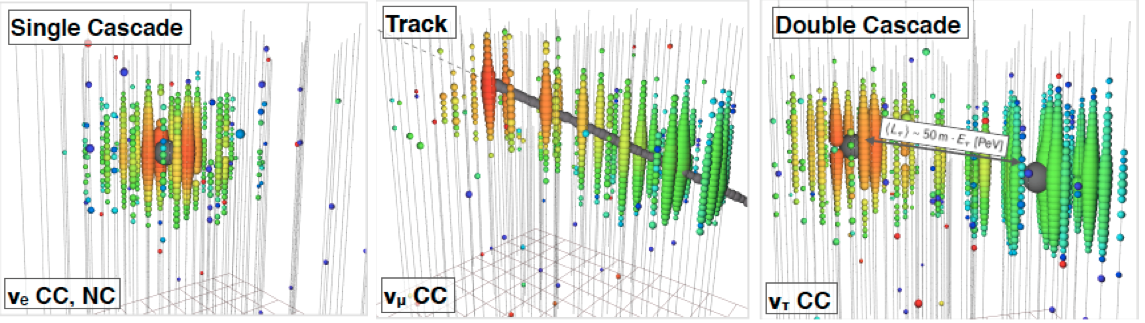}
	\caption{Different topologies of events that can be seen in large neutrino telescopes, such as IceCube, as a consequence of a $\nu_{e}$ CC or $\nu$ NC interactions (left panel), a $\nu_{\mu}$ interaction (middle panel), and a  $\nu_{\tau}$  interaction (right panel)~\cite{icrc-julianna}.\label{fig:topology}}
 \end{figure}

Another remarkable achievement has been the first detection of a neutrino tau candidate. Large neutrino telescopes are sensitive to neutrino flavour because the signal within the detector is flavour-dependent. Long and thin tracks are produced by relativistic muons crossing the detector, single cascades are created by electrons or neutral current interactions, and double cascades are signatures of the tau neutrino. A simulated double-cascade event, illustrated in the right panel of Fig.~\ref{fig:topology}, is produced by the interaction of the tau neutrino with the ice/water inside the detector. The track between the two cascades comes from the tau lepton subsequently traveling through the detector; the tau then decays to produce the second cascade.
 \begin{figure}[t!]
	\centering
	\includegraphics[width=0.5\textwidth]{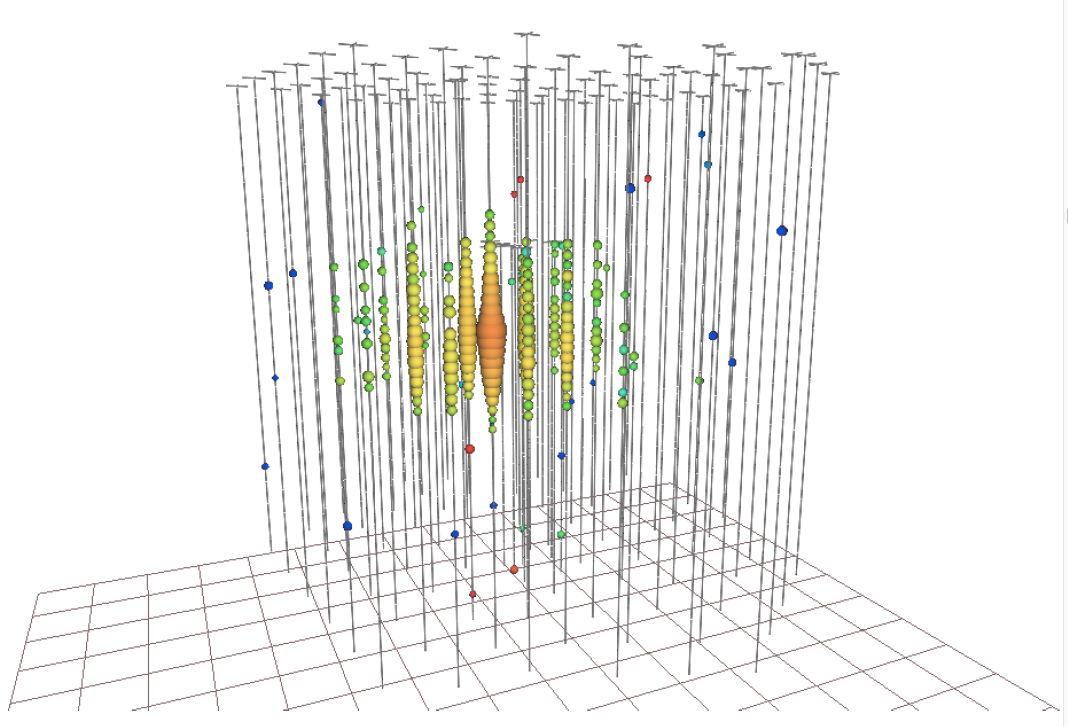}
	\caption{Event view of the first tau neutrino candidate seen in IceCube~\cite{icrc-dawn}.\label{fig:doubledouble}}
 \end{figure}
Several analyses, for example~\cite{icrc-julianna}, have been searching for the double cascade signature among the events detected by IceCube. One candidate was independently selected by the three analyses and constitutes strong evidence for the first detection of a high-energy tau neutrino~\cite{icrc-dawn,icrc-julianna}. This event, shown in Fig.~\ref{fig:doubledouble}, has a double cascade event topology with a deposited energy of 9 TeV in the first cascade and 80 TeV in the second, and a distance of 17 m between the cascades.
The particle identification of neutrino events (track, single or double cascade) allows an estimation of the flavor ratio of the HESE sample, leading to a best fit $\nu_e:\nu_{\mu}:\nu_{\tau}$ = 0.29 : 0.50 : 0.21~\cite{icrc-julianna}.
 
\section{What is next?}
In the next decade, the landscape of large neutrino telescopes is expected to drastically change with the deployment of the next generation of telescopes. 
Beyond the completion of the steadily growing GVD telescope~\cite{gvd-eps}, it is expected that by 2025 most of the KM3NeT ORCA and ARCA detectors as well as the IceCube-Upgrade will be deployed. 

KM3NeT, the next generation neutrino telescope, is currently under construction in the Mediterranean Sea~\cite{loi}. It is composed of two different nodes sharing the same detector technology: ARCA (Astroparticle Research with Cosmics in the Abyss) located off Capo Passero in Sicily and ORCA (Oscillation Research with Cosmics in the Abyss) located close to the ANTARES telescope off Toulon, France. While ARCA will be dedicated to searches for cosmic neutrino sources, ORCA has a denser spacing of optical modules and is optimized to detect GeV atmospheric neutrinos passing through the Earth in order to probe the neutrino mass hierarchy. At the time of writing, 4 ORCA and 1 ARCA strings are taking data. Each string is made of 18 Digital Optical Modules (DOM) with 9(36)~m average spacing in the ORCA (ARCA) configuration. To optimize the angular reconstruction and noise rejection, each KM3NeT DOM is equipped with 31 photomultipliers (PMT) sensitive to Cherenkov light. The construction will continue with higher speed in the coming years in order to reach full completion in 2025, with 115 strings on the ORCA site and 230 for ARCA. Details on the physics capabilities of ORCA and ARCA can be found in~\cite{icrc-rosa}.
We also note that besides being able to study neutrino physics and resolve the mass hierarchy, ORCA will be able to set competitive constraints on the GeV astrophysical neutrino flux~\cite{icrc-nmo}.

The IceCube-Upgrade, recently approved, will consist of  7 new strings to be deployed at the center of the detector inside the DeepCore region~\cite{icrc-aya}. The additional strings will be characterized by a reduced interstring spacing (20~m vs 70~m in DeepCore) and reduced inter-DOM spacing (2.4~m vs 7~m in DeepCore), as well as new sensors with increased photocathode coverage and ultraviolet sensitivity. This will improve the constraints we can set on atmospheric neutrino studies, allowing us for example to achieve a 10\% better sensitivity to $\nu_{\tau}$ normalization in one year of operations~\cite{icrc-dawn}. 
In addition to the DOMs, the Upgrade strings will include calibration devices to achieve a better measurement of the optical properties of the ice. These measurements will allow us to improve our calibration and archival analyses of the 10-year data collected by IceCube.
The deployment of the IceCube-Upgrade is planned for 2022-2023.

\section{Summary}
High-energy neutrino astronomy is a young field of research but it has already made several breakthroughs and collected many other noteworthy results. With the deployment of the next generation of large neutrino telescopes, GVD, KM3NeT, and the IceCube-Upgrade, even more results and discoveries are expected for the next decade.

\end{document}